\documentclass[namedreferences]{solarphysics}

\usepackage[hyperref,optionalrh,showbiblabels]{spr-sola-addons} 
\usepackage[optionalrh]{spr-sola-addons} 
\usepackage{epsfig}          
\usepackage{graphicx}        
\usepackage{natbib}         
\usepackage{amssymb}        
\usepackage[T1]{fontenc}        
\usepackage{url}             
\def\UrlFont{\sf}            

\chardef\us=`\_

\begin{document}

\begin{article}
\begin{opening}

\title{Analysis of Quiescent Corona X-ray Spectra from SphinX during the 2009 Solar Minimum}

\author[addressref={aff1},email={bs@cbk.pan.wroc.pl}]{\inits{B.}\fnm{B.}~\lnm{Sylwester}\orcid{0000-0001-8428-4626}} \sep
\author[addressref={aff1},email={js@cbk.pan.wroc.pl}]{\inits{J.}\fnm{J.}~\lnm{Sylwester}\orcid{0000-0002-8060-0043}} \sep
\author[addressref={aff1},email={ms@cbk.pan.wroc.pl}]{\inits{M.}\fnm{M.}~\lnm{Siarkowski}} \sep
\author[addressref={aff2},email={kennethjhphillips@yahoo.com}]{\inits{K.~J.~H.}\fnm{K.~J.~H.}~\lnm{Phillips}\orcid{0000-0002-3790-990X}} \sep
\author[addressref={aff1},email={pp@cbk.pan.wroc.pl}]{\inits{P.}\fnm{P.}~\lnm{Podgorski}} \sep
\author[addressref={aff1},email={mg@cbk.pan.wroc.pl}]{\inits{M.}\fnm{M.}~\lnm{Gryciuk}}



\address[id=aff1]{Space Research Centre, Polish Academy of Sciences (CBK PAN), Warsaw, Bartycka 18A, Poland}
\address[id=aff2]{Scientific Associate, Earth Sciences Dept., Natural History Museum, Cromwell Road, London SW7 5BD, UK}

\runningauthor{B. Sylwester {\em et al.}}
\runningtitle{SphinX Spectra of Quiescent Corona}

\begin{abstract}
The SphinX X-ray spectrophotometer on the {\em CORONAS-PHOTON}\/ mission observed the 1~--~15\,keV X-ray spectrum of the spatially integrated solar corona during the deep minimum of 2009, when solar activity was exceptionally low. Its sensitivity for energies $> 1.2$\,keV was higher than that of any other solar X-ray spectrometer in orbit at the time, including the detectors on GOES. Using much improved instrumental data than was used previously, we analyzed SphinX spectra in 576 intervals for which there was no discernible activity (NA), 40 intervals when there were  X-ray brightenings (B), and sixteen intervals when there were micro-flares with peak emission less than GOES A1 (F). An instrumental background spectrum, formed over 34 hours of spacecraft night-time periods and including electronic noise and particle radiation, was subtracted from the solar spectra. Theoretical spectra were used to deduce temperatures on an isothermal assumption for the NA, B, and F intervals (1.69, 1.81, and 1.86\,MK respectively). Differential emission measure (DEM) analysis for the same spectra revealed a ``cooler'' component (log\,$T = 6.2$ or $T \approx 1.6$\,MK) in each case, but with a second hotter component having a less well-defined peak temperature varying from $\approx 2.5$ to $\approx 3.5$\,MK (log\,$T = 6.4$ and 6.55) and an emission measure between two and three orders smaller than that of the cooler component. These results are similar to those obtained at times just after solar minimum with the EVE instrument. A very hot component that might indicate the signature of nano-flare heating of the corona is not evident in SphinX data.
\end{abstract}
\end{opening}


\section{Introduction}\label{sec:Intro}


The unusually low levels of solar activity during the 2009 solar minimum between Cycles 23 and 24 has been remarked on by several authors ({\em e.g.,} \cite{she10}). The soft X-ray emission frequently dropped below the GOES A1 level (equivalent to $10^{-8}$~W~m$^{-2}$~s$^{-1}$) for days or weeks at a time, so that accurate monitoring solar X-ray emission in the usual GOES bands (1~--~8~\AA, 0.5~--~4~\AA) was not possible at times. However, the Polish SphinX instrument ({\em Solar Photometer IN X-rays}), operating in similar energy range to the GOES 1~--~8\,\AA\ band, was able to continue the monitoring over most of the 2009 minimum period.

SphinX was a highly sensitive X-ray spectrophotometer included as part of the TESIS instrument package \citep{kuz09} that flew on the Russian {\em CORONAS-PHOTON}\/ solar mission. The spacecraft operated throughout most of the 2009 period, with SphinX measurements available from 20~February to 28~November. Full instrumental details were given by \cite{jsyl08_Sph} and \cite{jsyl12_Sph}. Briefly, SphinX had a total of four PIN diode detectors (D1~--~D4), each consisting of silicon wafers with beryllium windows. The most significant emission was from detector D1, the energy range of which was 1.2~--~15~keV (0.8~--~10~\AA), and energy resolution (FWHM) of 464~eV. The instrument was intensity-calibrated to better than 5\,\% accuracy using ground-based facilities before launch. The spacecraft's near-polar orbit enabled SphinX measurements to be made in sixty or more minutes of the 96-minute orbital period, with continuous coverage for two-week periods in April and July.

Here we report on a detailed analysis of time intervals in the SphinX data set, comparing SphinX spectra in some 576 time intervals during which solar activity was exceptionally low (designated NA, or ``no activity'') and 40 time intervals when there were minor brightenings (B), in order to study the nature of the quiet-Sun X-ray-emitting corona during the exceptionally deep minimum of 2009. In addition, some sixteen intervals were identified by eye as having flare-like increases in X-ray emission that we refer to as micro-flares (F). The averaged spectra in the NA, B, and F intervals were analyzed first on an isothermal assumption and then evaluating differential emission measure (DEM) using methods given in previous work by us. The isothermal analysis is found not to give a satisfactory fit to any of the spectra, particularly at higher energies ($\gtrsim 2.5$\,keV). The DEM analysis on the other hand gives an improved fit to the observed spectra and suggests the presence of both a low-temperature (``cooler'') and a higher-temperature (``hotter'') component, the latter having for all cases a peak DEM of between two and three orders less than that of the cooler component.


\section{Observed X-ray Emission as Recorded by SphinX}\label{sec:Sph_X-ray_Em}

\begin{figure}
   \centerline{\includegraphics[width=12cm]{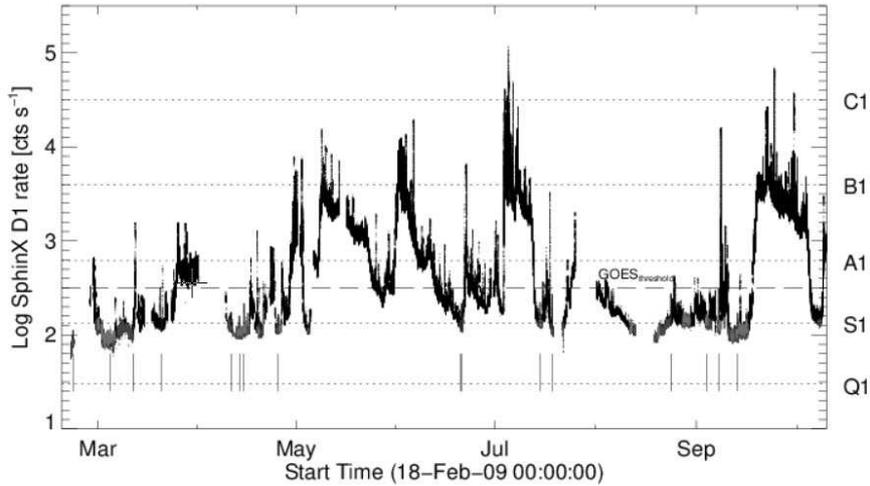}}
      \caption{The X-ray emission as recorded by the SphinX D1 detector in photon counts s$^{-1}$, plotted logarithmically over the period 20~February to 9~October 2009 (at later times solar activity increased and data were excluded for this study). Red portions indicate the times of the 576 NA intervals when the count rate was $< 140$\,s$^{-1}$ and green portions times of 40 small brightenings (B). Times of the sixteen micro-flares are indicated by thin vertical blue lines beneath the light curve (intervals F: see Table~\ref{tab:sixteen_flares} and Figure~\ref{fig:lightcurves_6_flares}). The GOES 1~--~8\,\AA\ lower threshold is indicated by the horizontal dashed line, and the GOES A1, B1, C1 levels and the S1 and Q1 levels (see text) are indicated by horizontal dotted lines. }
      \label{fig:D1_lc}
  \end{figure}

In Figure~\ref{fig:D1_lc}, the X-ray light curve as recorded by detector D1 (photon count rate, s$^{-1}$) is shown as a logarithmic plot over the period 20~February to 9~October 2009 (at later times, solar activity increased and data were not used for the present study). The data analyzed here were selected from times outside intervals when there were high background, {\em e.g.} passages through the South Atlantic Anomaly or auroral ovals. The NA intervals (indicated in red) denote periods of at least five minutes when the D1 count rate was less than 140~s$^{-1}$ and with insignificant changes. Brightenings (B, indicated in green) are defined to be periods when there was an increase in count rate, starting from a count rate level $< 140$~s$^{-1}$, with a more or less symmetrical time profile. The microflares (F, times of which are indicated by short blue vertical lines beneath the light curve), identified by eye, were defined to be increases starting from count rates $< 140$~s$^{-1}$ but with a sharper rise and slower decline, like those of fully fledged flares. Generally, B events had a smaller amplitude than microflares (F). Note that a NA period may occasionally have a count rate higher than a B period. (In previous work \citep{jsyl12_Sph}, much longer intervals~--~2.2 to 18.2\,hours~--~were chosen.) The GOES 1~--~8\,\AA\ detection threshold corresponding to the SphinX detector D1 count rate is indicated in Figure~\ref{fig:D1_lc}, based on a cross-calibration analysis by \cite{gbu13}, as are fluxes in the 1~--~8\,\AA\ range corresponding to GOES A1, B1, and C1 levels; we label two additional levels, S (for ``small''), corresponding to $10^{-9} - 10^{-8}$\,W\,m$^{-2}$, and Q (for ``quiet''), corresponding to $10^{-10} - 10^{-9}$\,W\,m$^{-2}$ \citep{gbu13}. The S1 ($10^{-9}$\,W\,m$^{-2}$) level is a factor 10 lower than the GOES 1~--~8\,\AA\ A1 ($10^{-8}$\,W\,m$^{-2}$) level, the Q1 ($10^{-10}$\,W\,m$^{-2}$) level a factor 100 lower.

\section{Spectral Analysis}\label{sec:Sp_anal}

\begin{figure}
   \centerline{\includegraphics[width=12cm]{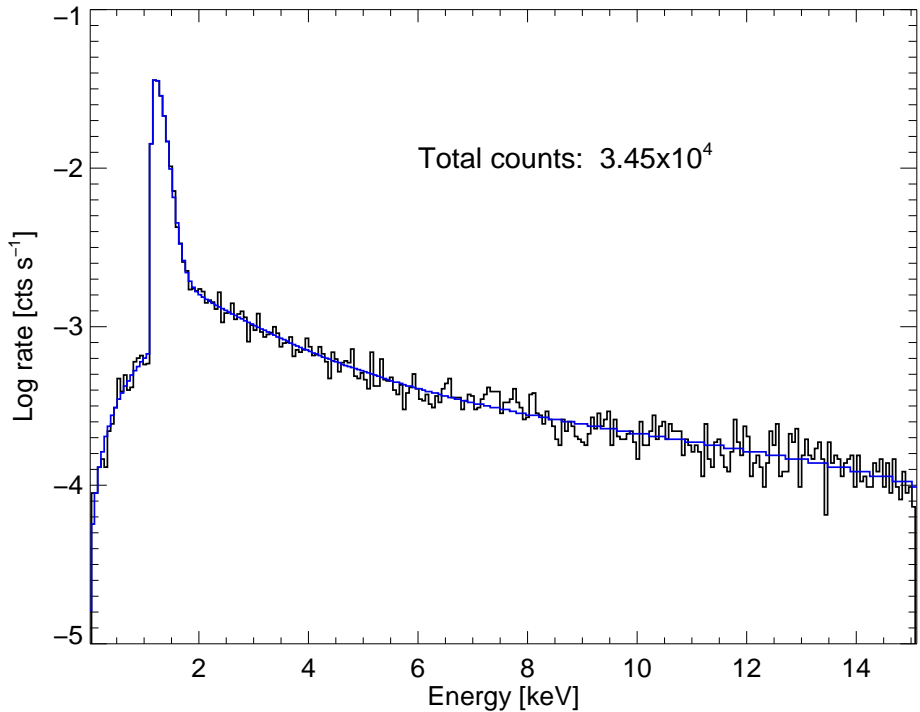}}
      \caption{Logarithmic plot of count rate per spectral bin against photon energy (keV) of the background X-ray spectrum of SphinX D1 detector (counts s$^{-1}$: black histogram) as recorded in approximately 34~hours of spacecraft night-time (total photon counts indicated). The continuous blue curve is a best-fit approximation to the histogram which was subtracted from solar spectra observed by SphinX. }
      \label{fig:nighttime_spectrum}
  \end{figure}

Spectra during the NA, B, and F intervals were derived and first analyzed using an isothermal assumption. A notable difference in the present analysis from our previous work is the use of a much improved background spectrum. This spectrum, shown in Figure~\ref{fig:nighttime_spectrum}, was obtained by summing a total of approximately 34 hours of spacecraft night-time observations, taken from regions across the Earth that avoided spacecraft passages through the South Atlantic Anomaly and polar auroral regions, and times outside of magnetic substorms. Its constancy with time was confirmed for the first two months of the mission (data at later times were unavailable to conserve the spacecraft data quota assigned to SphinX). The background includes electronic noise (mostly at lower energies, $\lesssim 2$\,keV), particle emission (present at all energies and due to the particle environment of the spacecraft), and fluorescence from aluminium which makes up the bulk of the instrument structure (energies $\lesssim 2$\,keV). The shape of this background spectrum was approximated by a continuous curve (shown in the figure), and all solar spectra taken by SphinX were corrected by subtracting this measured background. Note that, even at low levels of solar activity considered here, the background emission amounted to $< 1$\% of the solar X-ray emission.

\begin{figure}
   \centerline{\includegraphics[width=12cm]{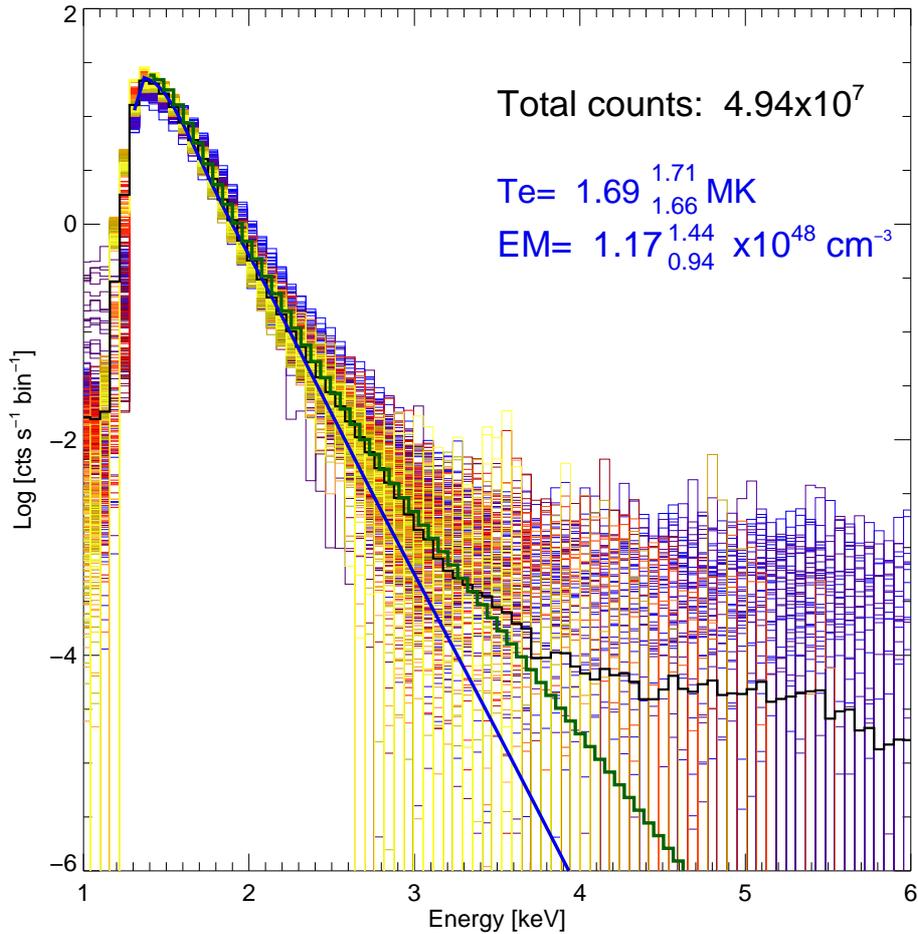}}
      \caption{SphinX NA spectra in the 1~--~6\,keV range (thin-line histograms) for the 576 time intervals when there was no discernible solar activity and their average (thick black histogram). The best-fit isothermal spectrum is shown by the thick blue curve (temperature and emission measure values shown in the figure) and the thick green-line histogram shows the fit by folding the DEM solution (Section~\ref{sec:DEM_anal}) through the SphinX spectral response function. The background spectrum in Figure~\ref{fig:nighttime_spectrum} has been subtracted for all spectra shown. A rough distinction in the times of the spectra is indicated by the colours of the histogram (blue = early in the mission, red = middle of the mission, yellow = late in the mission). }
      \label{fig:aver_spectrum_576_MS}
  \end{figure}

\begin{figure}
\centerline{\hspace*{0.005\textwidth}
             \includegraphics[width=0.49\textwidth,clip=,angle=0]{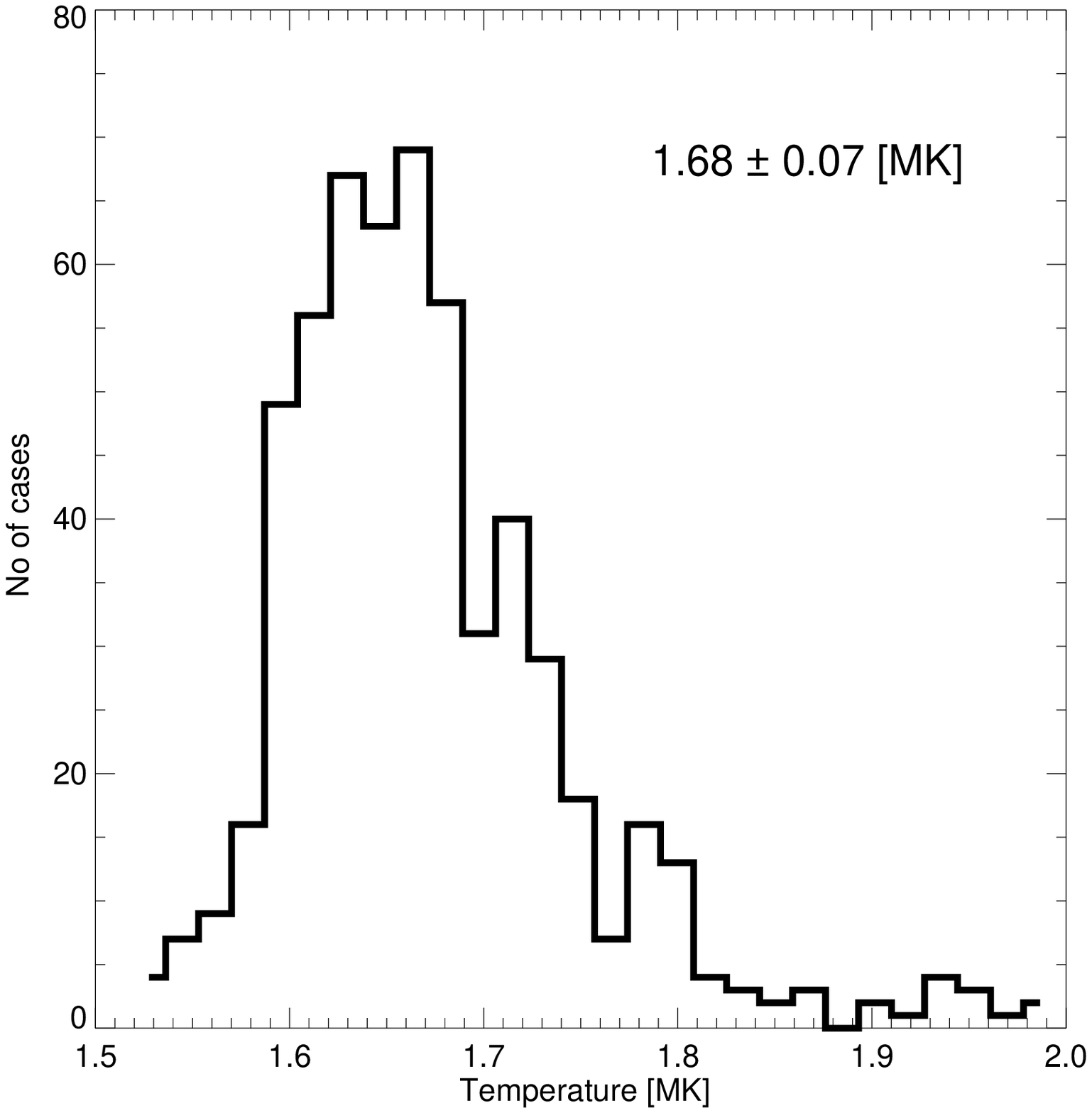}
              \hspace*{0.01\textwidth}
             \includegraphics[width=0.49\textwidth,clip=]{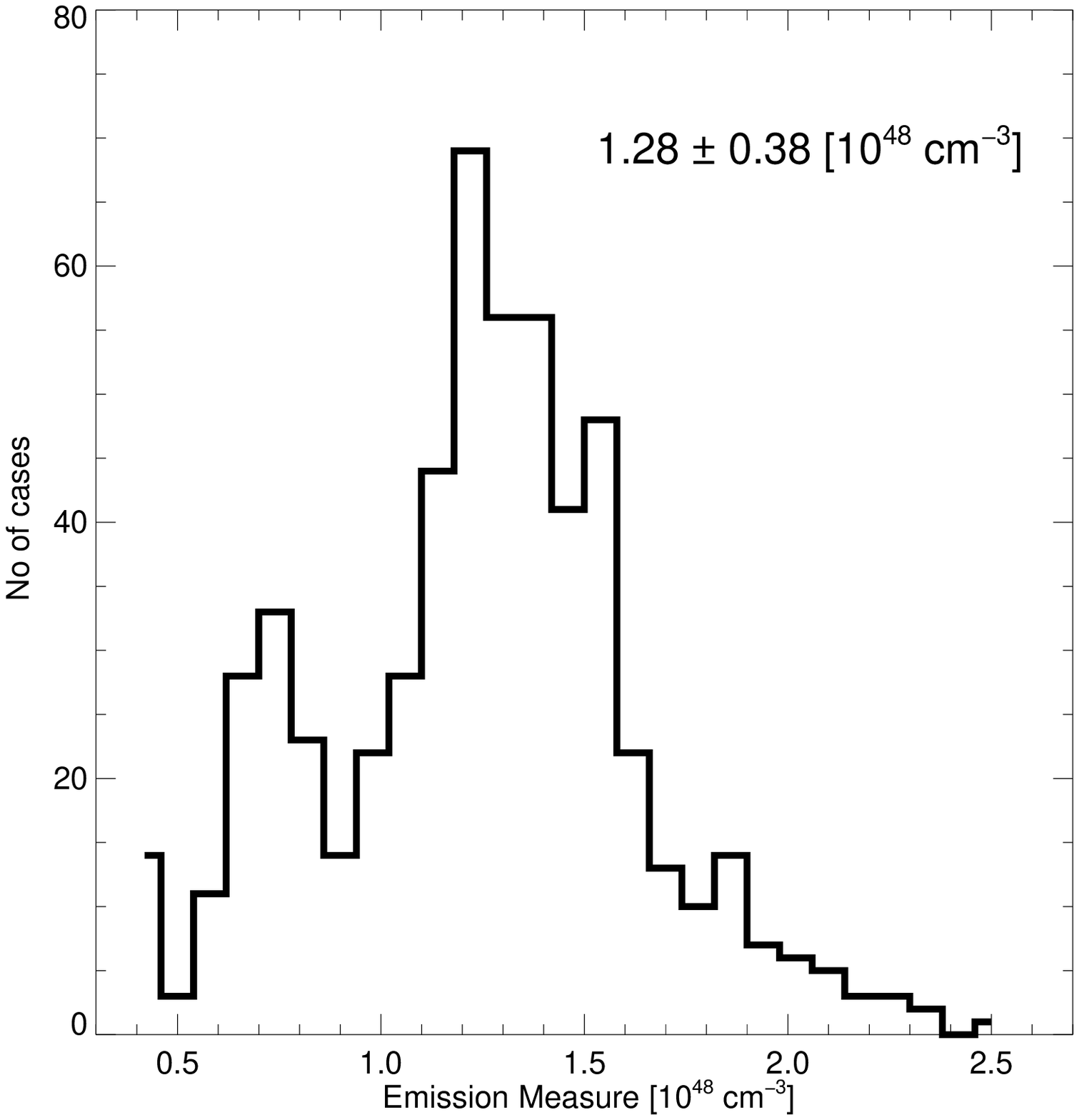}}
             \label{hist_T_EM}
             \caption{Distribution of temperatures (left panel) and emission measures (right panel) for isothermal fits to 576 NA spectra.}
             \label{fig:TandEM_for_NA_sp}
\end{figure}

Individual spectra for all 576 NA time intervals are shown in Figure~\ref{fig:aver_spectrum_576_MS}, colour-coded according to time of registration (blue = early in the mission; red = middle of the mission; yellow = late in the mission) together with an isothermal fit to the averaged spectrum (blue curve in the figure). The isothermal fit was derived by folding through the SphinX spectral response matrix theoretical spectra from the  CHIANTI (v.~9) atomic database and software package \citep{der97,der19} and selecting the fit with the least $\chi^2$. Use of CHIANTI requires choice of ionization fractions (we took the default values in CHIANTI) and abundances (mostly based on \cite{fel92b} but with values for Si, S, Cl, Ar, and K from RESIK X-ray spectrometer analyses: see \cite{bsyl11,bsyl15}). The temperature and emission measure from the averaged NA spectrum, based on $4.94 \times 10^7$ photon counts, were found to be $1.69 \pm 0.02$\,MK and $1.17 ^{1.44}_{0.94} \times 10^{48}$\,cm$^{-3}$ respectively. A previous analysis \citep{jsyl12_Sph}, based on a low-activity interval of just under six hours on 16~September and with previous software versions and the \cite{fel92b} abundances, gave temperature and emission measure values, $1.71 \pm 0.02$\,MK and $9.78 \pm 2.0 \times 10^{47}$\,cm$^{-3}$, which are consistent within the error ranges. Figure~\ref{fig:TandEM_for_NA_sp} shows the distribution of temperatures and emission measures for NA spectra, illustrating the comparative narrowness of the temperature and emission measure ranges. It is clear that the isothermal fit in Figure~\ref{fig:aver_spectrum_576_MS} falls well below most of the observed NA spectra at energies $\gtrsim 2.5$\,keV and so a more detailed analysis involving DEMs is required, which is dealt with in Section~\ref{sec:DEM_anal}.

\begin{figure}
   \centerline{\includegraphics[width=12cm]{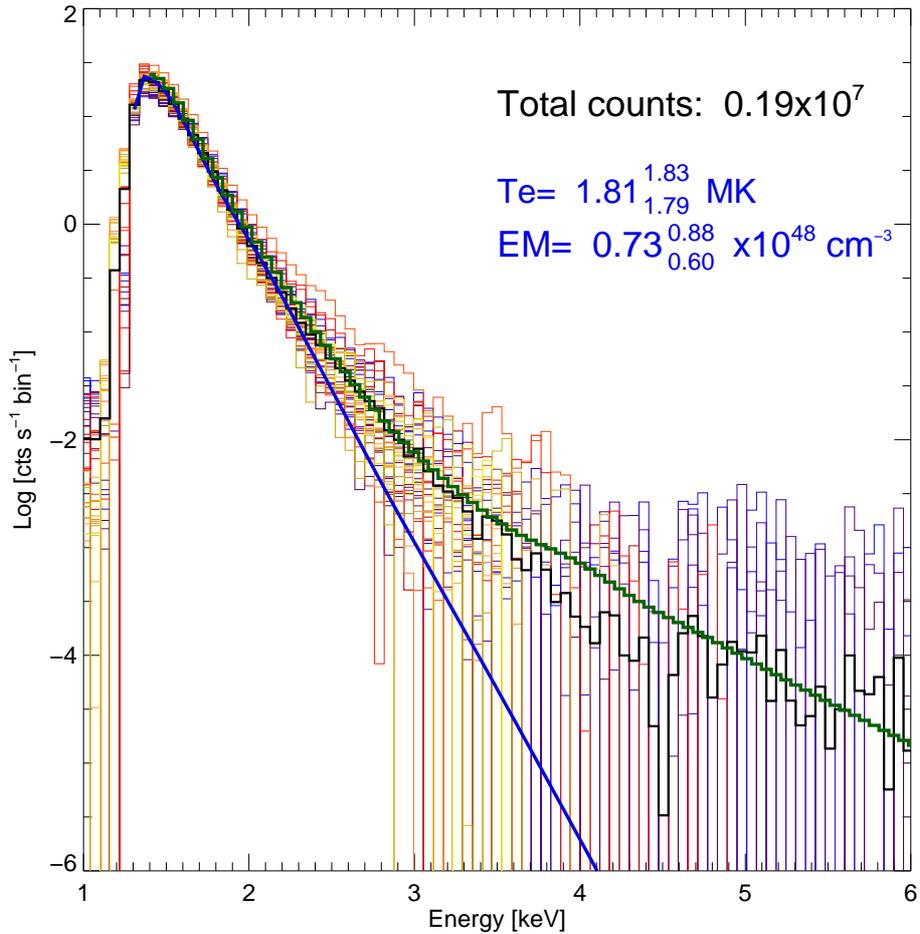}}
      \caption{SphinX B spectra in the 1~--~6\,keV range (thin-line histograms) with averaged spectrum (thick black histogram), and best-fit isothermal spectrum (thick blue curve: temperature and emission measure values indicated in the figure) for the 40 time intervals when there were tiny brightenings. The thick-line green histogram shows the fit to the averaged spectrum by folding the DEM solution (Section~\ref{sec:DEM_anal}) through the SphinX spectral response function. The background spectrum (Figure~\ref{fig:nighttime_spectrum}) has been subtracted for all spectra shown. Histograms colours are a rough indication of time (blue = early, red = middle, yellow = late in the mission). }
      \label{fig:aver_spectrum_bri_MS}
  \end{figure}

The isothermal analysis was also applied to the 40 SphinX detector D1 spectra when there were small brightenings (B). Individual and averaged B spectra are shown in Figure~\ref{fig:aver_spectrum_bri_MS} together with the best isothermal fit to the averaged spectrum. For this case, the temperature and emission measure were $1.81 \pm 0.02$\,MK and $0.73 ^{0.88} _{0.60} \times 10^{48}$\,cm$^{-3}$, so the temperature is slightly higher and the emission measure lower than the quiet-Sun (NA) case. As with the NA spectral fit in Figure~\ref{fig:aver_spectrum_576_MS}, the isothermal fit diverges from the observed spectra for energies $\gtrsim 2.5$\,keV, so indicating the need for a DEM analysis, as is done in Section~\ref{sec:DEM_anal}.

\begin{figure}
   \centerline{\includegraphics[width=12cm]{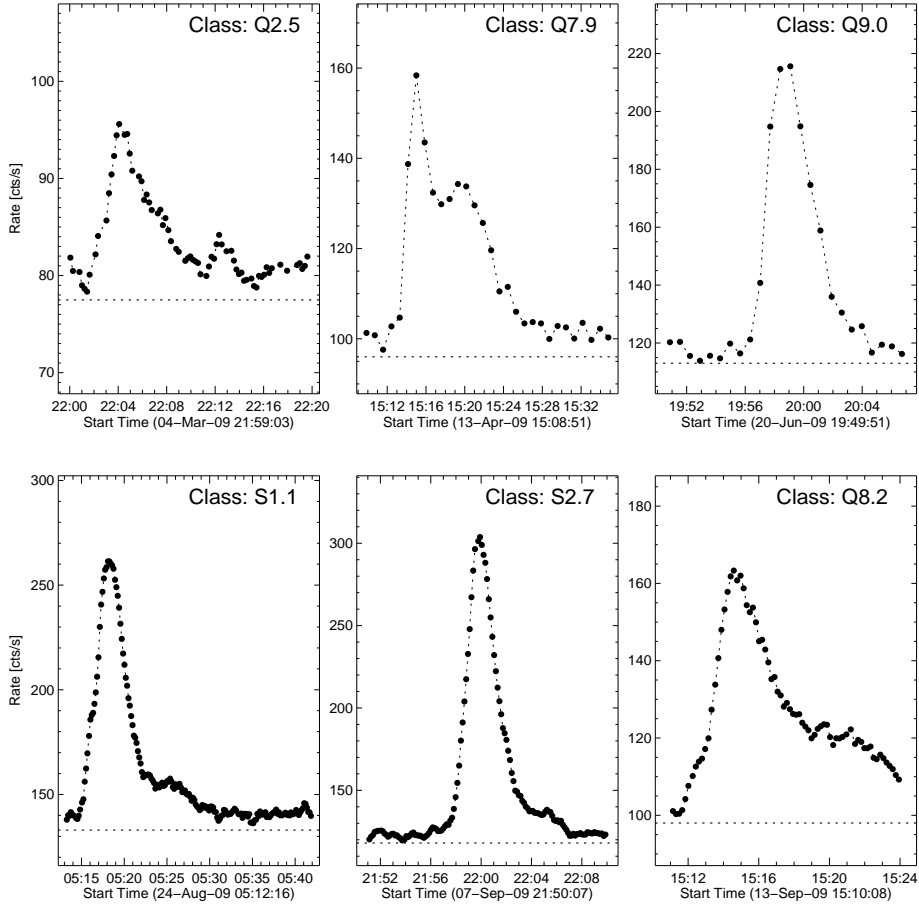}}
      \caption{SphinX light curves of the six of the sixteen micro-flares analyzed here. The micro-flare peak emission above a pedestal level, marked as the dotted line in each panel, is indicated, using the definitions of Section~\ref{sec:Sph_X-ray_Em}. The D1 total count rate (s$^{-1}$) in each case is plotted linearly against time (UT).}
      \label{fig:lightcurves_6_flares}
  \end{figure}

Data for the sixteen microflares (F) identified in low-activity intervals are given in Table~\ref{tab:sixteen_flares}, while SphinX detector D1 X-ray light curves for six of the sixteen micro-flares are shown in Figure~\ref{fig:lightcurves_6_flares}. The time variations are essentially the same as more developed flares (see, {\em e.g.}, the SphinX flares analyzed by \cite{gry17}), with rise times of one to four minutes and longer decay times, up to seven minutes. In terms of the emission classes mentioned in Section~\ref{sec:Sph_X-ray_Em}, the flares have peak emission above a pre-flare pedestal level (marked in Figure~\ref{fig:lightcurves_6_flares} as a dotted line) varying from Q0.3 ($3 \times 10^{-11}$ W~m$^{-2}$) to S1.6 ($1.6 \times 10^{-9}$ W~m$^{-2}$), and so are far below the GOES A1 level.

\begin{figure}
   \centerline{\includegraphics[width=12cm]{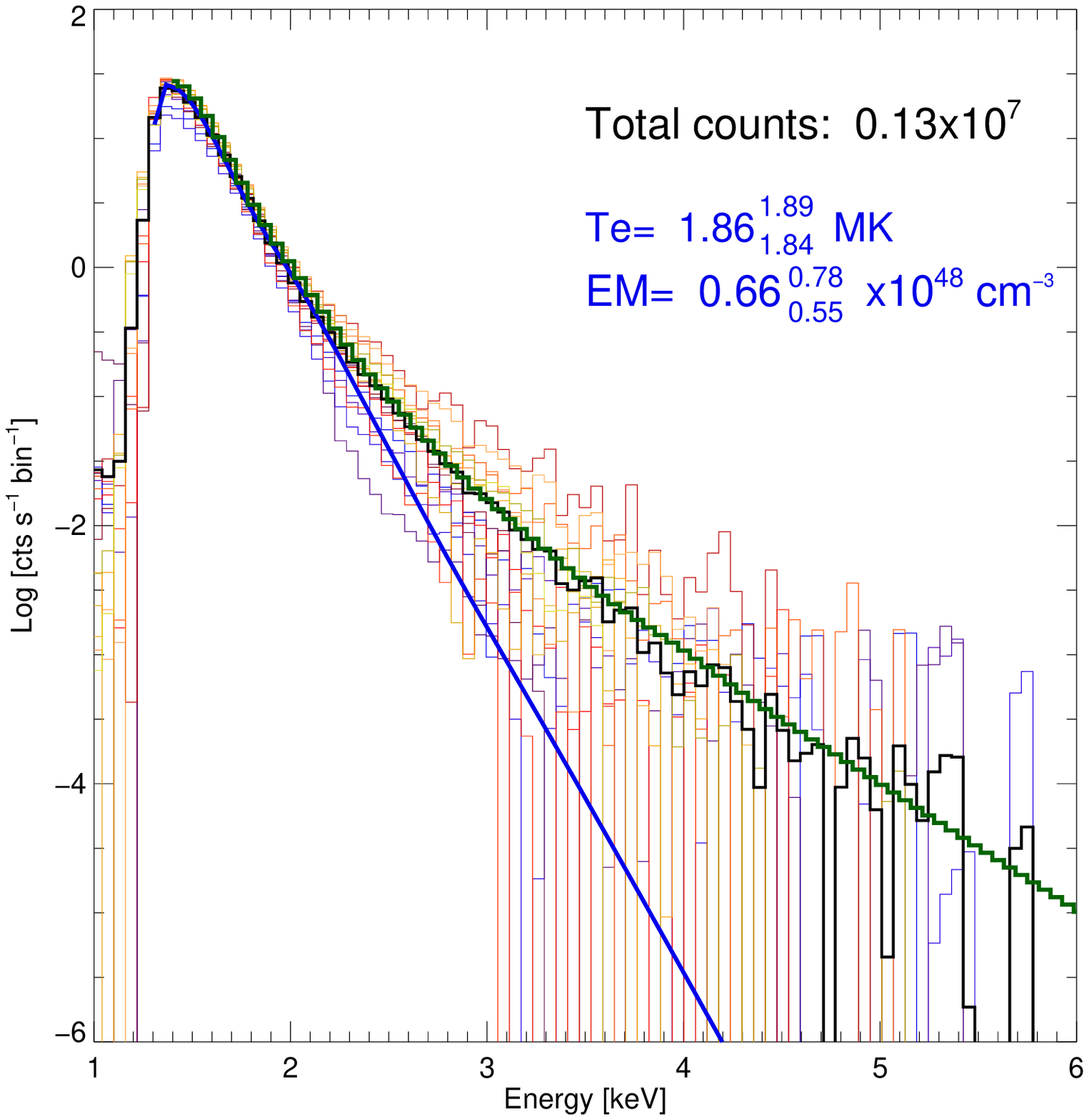}}
      \caption{SphinX F spectra in the 1~--~6\,keV range (thin-line histograms) with averaged spectrum (thick black histogram), and best-fit isothermal spectrum (thick blue curve: temperature and emission measure values indicated in the figure) for the sixteen time intervals when there were micro-flares. The histogram colours roughly distinguish times (blue = early, red = middle, yellow = late in the mission). The thick-line green histogram shows the fit to the averaged spectrum by folding the DEM solution (Section~\ref{sec:DEM_anal}) through the SphinX spectral response function. The background spectrum (Figure~\ref{fig:nighttime_spectrum}) has been subtracted for all spectra shown, but the spectra include the pre-flare pedestal. }
      \label{fig:aver_spectrum_fla_MS}
  \end{figure}

In Figure~\ref{fig:aver_spectrum_fla_MS} we show the SphinX spectra averaged over intervals of a few seconds around the peak of each microflare (thin line histograms), with different colours as an indication of the microflare times. The spectra are background-subtracted but include the pre-flare pedestal level. The thick-line histogram is the average spectrum of these sixteen spectra, and includes a total of $1.3 \times 10^6$ photon counts. The best-fit isothermal spectrum (using CHIANTI theoretical spectra as before) shown in the figure has temperature and emission measure equal to $1.86 \pm 0.02$\,MK and $0.66 ^{0.78}_{0.55} \times 10^{48}$\,cm$^{-3}$. As with the NA and B spectra, the isothermal fit diverges from most of the individual spectra at energies higher than $\approx 2.2$~keV, so indicating the necessity for a differential emission measure analysis (Section~\ref{sec:DEM_anal}).

Because of the high sensitivity of SphinX to $1 - 8$\,\AA\ emission, the microflares selected here are far less intense than those in a comprehensive study of active region microflares seen with the {\em Reuven Ramaty High Energy Solar Spectroscopic Imager}\/ (RHESSI: \cite{han08a}). To illustrate this point, we made a rough estimate of the total thermal energy $E_{\rm th} = 3 N_{\rm e} k_{\rm B} T_{\rm e} V$ ($N_{\rm e} = $ electron density, $T_{\rm e} = $ electron temperature, $k_{\rm B} = $ Boltzmann's constant, $V = $ emitting volume) in the first of the microflares listed in Table~\ref{tab:sixteen_flares} (21~February) using the temperature $T_{\rm e}$ and volume emission measure EM$ = N_{\rm e}^2 V$ from the isothermal fit to the SphinX spectrum at 11:40~UT and an image of this microflare seen by the {\em Sun Earth Connection Coronal and Heliospheric Investigation}\/ (SECCHI) (171\,\AA\ filter) on the STEREO-B spacecraft. The SphinX observations indicate a temperature $T = 1.65$\,MK and EM$ = 1.0 \times 10^{47}$\,cm$^{-3}$ with a pre-flare level subtracted. The SECCHI images of this microflare allows us to estimate a volume of $3.8 \times 10^{24}$\,cm$^3$, so combining this with the SphinX emission measure an electron density $N_{\rm e} = 1.6\times 10^{11}$\,cm$^{-3}$ is implied assuming a unit filling factor $f=1$. The total thermal energy is then $E_{\rm th} = 4.2\times 10^{26}$\,erg, with correspondingly higher values for smaller filling factors $f$. This compares with a median thermal energy of $10^{28}$ erg from the RHESSI active region flares of \cite{han08a}, a value that is comparable with the combined SphinX and SECCHI value only if $f \approx 10^{-2}$. Although very small filling factors are not unknown, it would appear likely that the microflares seen with SphinX are generally much smaller than the active region flares of \cite{han08a}. The very low sensitivity of SphinX to high energies ($\gtrsim 20$~keV) does not enable us to comment on the presence or absence of a non-thermal peak as has been seen with the RHESSI microflares.

\begin{table}
\caption{Characteristics of sixteen microflares seen by SphinX.}
\label{tab:sixteen_flares}
\begin{tabular}{llcccrr}
\hline                   
Date (2009) & Peak time  & Peak X-ray & Peak $T$ & Peak EM & $t_{\rm rise}$  & $t_{\rm fall}$  \\
& (UT) & flux class\tabnote{Pre-flare emission subtracted.} & (MK) & $10^{47}$ cm$^{-3}$ & (s) & (s) \\
  \hline
21 Feb. & 11:40:23 & Q0.6 & 1.65 & 1.0 & 12  & 38 \\
04 Mar. & 22:04:00 & Q0.4 & 1.65 & 0.9 &  75 & 190 \\
11 Mar. & 21:34:00 & Q0.3 & 1.68 & 0.5 &  22 & 90 \\
20 Mar. & 17:43:00 & Q0.9 & 1.73 & 0.7 & 160 & 440 \\
11 Apr. & 01:52:00 & Q2.2 & 1.69 & 1.3 &  75 & 165 \\
13 Apr. & 15:15:00 & Q2.9 & 1.68 & 1.9 &  60 & 105 \\
14 Apr. & 23:25:40 & Q7.3 & 1.71 & 2.6 &  64 & 128 \\
25 Apr. & 12:52:00 & Q1.9 & 1.74 & 1.1 & 220 & 340 \\
20 June  & 15:22:00 & Q0.3 & 1.76 & 0.4 &  90 & 310 \\
20 June  & 19:58:40 & Q6.8 & 1.73 & 2.7 &  60 & 100 \\
14 July  & 21:12:20 & Q3.3 & 1.69 & 1.8 &  70 & 185 \\
18 July  & 17:35:00 & S1.3 & 1.70 & 3.5 &  70 & 160 \\
24 Aug.& 05:18:10 & S1.0 & 1.94 & 1.6 &  75 & 120 \\
04 Sept. & 06:28:20 & Q2.0 & 1.97 & 0.6 &  10 &  37 \\
07 Sept. & 21:59:35 & S1.6 & 1.99 & 1.8 &  45 &  95 \\
13 Sept. & 15:14:35 & Q3.5 & 1.95 & 1.0 &  60 & 125 \\

 \\
  \hline
\end{tabular}
\end{table}

\section{Differential Emission Measure Analysis}\label{sec:DEM_anal}

As can be seen from Figures~\ref{fig:aver_spectrum_576_MS}, \ref{fig:aver_spectrum_bri_MS}, and \ref{fig:aver_spectrum_fla_MS}, the observed spectra diverge significantly from the isothermal curves for energies higher than about 2.5\,keV, the presence of a higher-temperature component being implied so necessitating the need for a differential emission measure analysis. Following our DEM analyses in previous work \citep{bsyl15}, we used an iterative procedure (called here the ``Withbroe--Sylwester'' method) based on a Bayesian technique and the work of \cite{with75} and \cite{jsyl80} and again using CHIANTI theoretical spectra. The solutions were obtained over a temperature range 1.3\,MK to 7\,MK (log\,$T = 6.11$ to 6.85, $T$ in K).

\begin{figure}
   \centerline{\includegraphics[width=12cm]{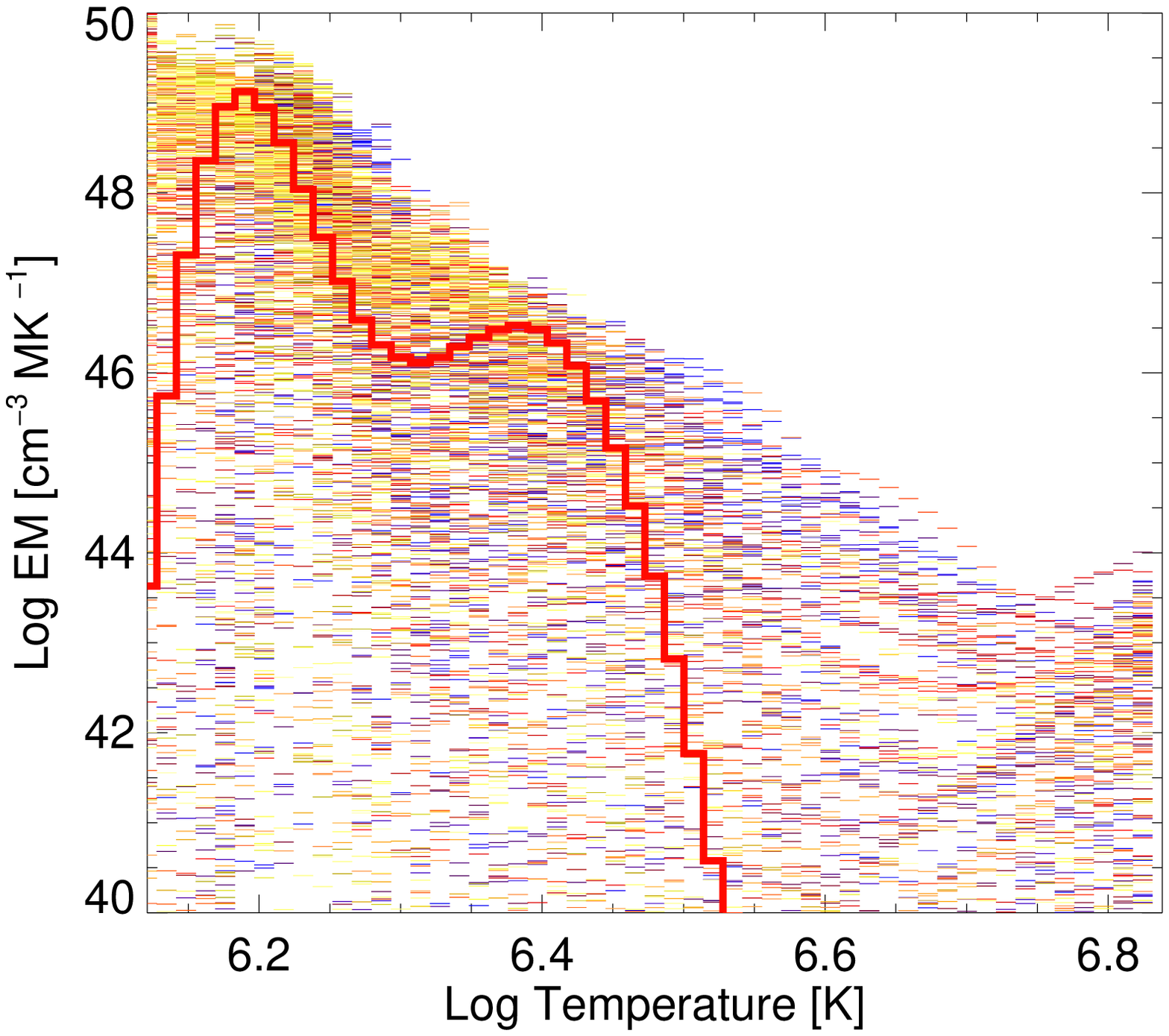}}
      \caption{Logarithm of differential emission measures (cm$^{-3}$~MK$^{-1}$) plotted against the logarithm of temperature (in K) derived for each of the 576 individual NA spectra (shown in various colours), together with the DEM calculated for the average of the 576 spectra (thick red histogram). }
      \label{fig:DEM_for_ind_spectra_1pt3_7_MK}
  \end{figure}

The resulting DEM for the 576 individual NA spectra and the averaged NA spectrum in Figure~\ref{fig:aver_spectrum_576_MS} is shown in Figure~\ref{fig:DEM_for_ind_spectra_1pt3_7_MK}, plotted against log\,$T$. The DEM solutions for the individual NA spectra are indicated by the scatter of histogram points in each interval (0.02) of log\,$T$, while the thick red histogram is the DEM for the averaged NA spectrum. A lower-temperature or cooler component is apparent with log\,$T = 6.18$ (1.5\,MK) as well as a hotter component having a peak temperature given by log\,$T = 6.38$ (2.4\,MK), but with peak DEM nearly three orders of magnitude smaller than that of the cooler component. The cooler component has a temperature that is very similar to that derived from the isothermal fit ($1.69 \pm 0.02$\,MK) shown in Figure~\ref{fig:aver_spectrum_576_MS} arising from the fact that the fits are dominated by the much larger spectral fluxes at lower ($\lesssim 2$\,keV) energies.

The DEM distribution for the averaged NA spectra was then folded through the instrument response function to give a spectrum (green histogram) that may now be compared with the observed spectra and isothermal fit shown in Figure~\ref{fig:aver_spectrum_576_MS}. It can be seen that the addition of the hotter component, even though it has a much smaller peak DEM, results in an improvement in fitting the observed spectra and their average at energies $\gtrsim 2.5$\,keV up to $\approx 3.5$\,keV. The addition of the hotter component is evidently needed to accomplish this improvement, so this analysis indicates that the quiet Sun X-ray-emitting corona, even at the extremely low activity levels during 2009, is non-isothermal, having temperature components of approximately 1.5\,MK and 2.4\,MK.

\begin{figure}
   \centerline{\includegraphics[width=12cm]{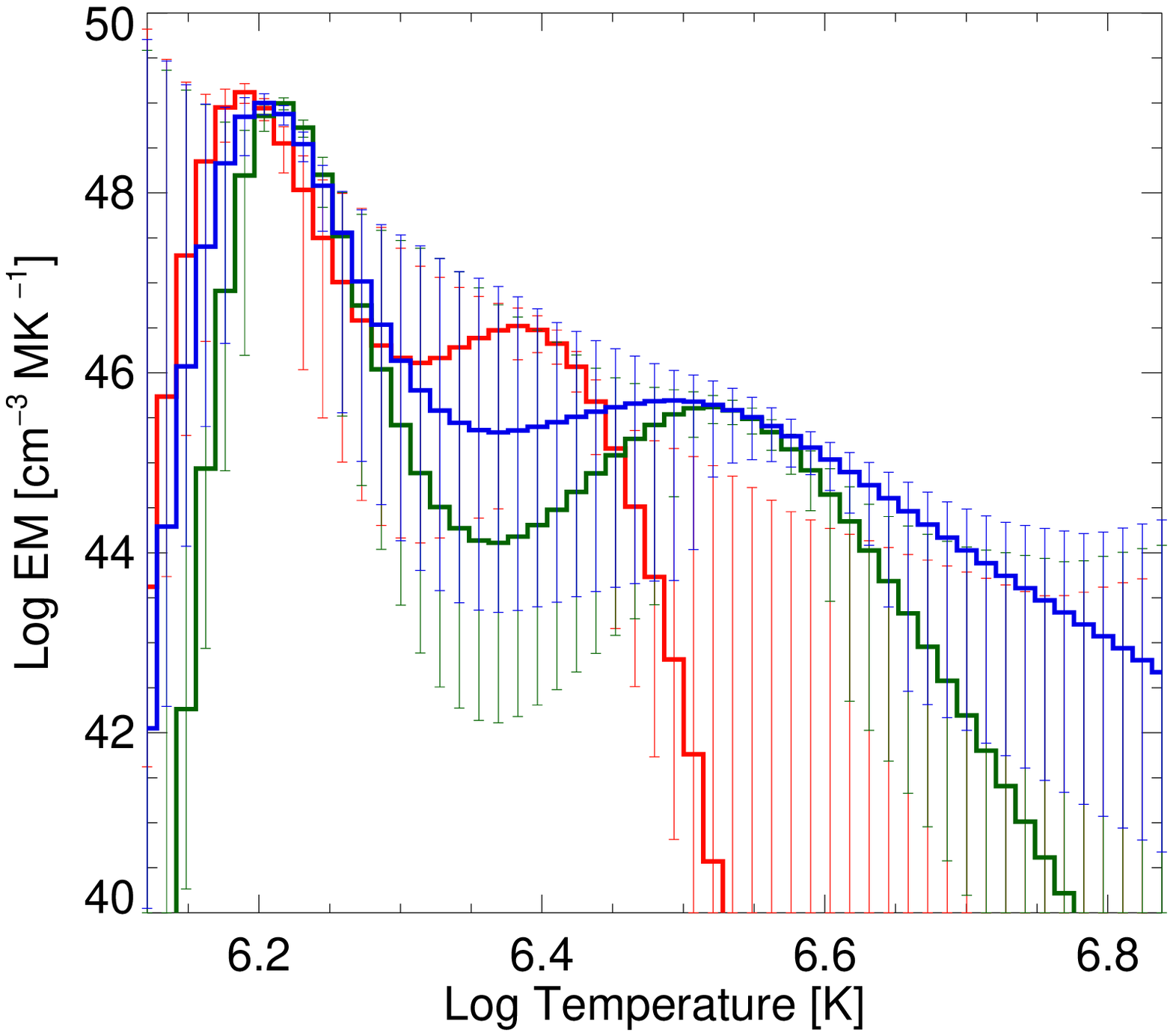}}
      \caption{Logarithm of differential emission measures (cm$^{-3}$~MK$^{-1}$) plotted against the logarithm of temperature (in K) for the average spectra in the NA (red), B (green), and F (blue) intervals. The uncertainties in the DEM solutions are indicated by the error bars in appropriate colours. }
      \label{fig:Dem_quiet_bri_flare}
  \end{figure}

The corresponding DEM distributions using the Withbroe--Sylwester procedure for the averaged B (green histogram) and F (blue histogram) intervals are shown in Figure~\ref{fig:Dem_quiet_bri_flare} with the DEM for the NA intervals (histogram with the thick red line in Figure~\ref{fig:DEM_for_ind_spectra_1pt3_7_MK}). Estimated uncertainties are indicated by the error bars with appropriate colours. For the B spectra shown in Figure~\ref{fig:aver_spectrum_bri_MS}, folding the DEM of Figure~\ref{fig:Dem_quiet_bri_flare} through the spectral response function results in a considerably improved fit (green histogram) to the averaged spectrum up to the observational limit ($\approx 6$\,keV). The cooler component in this case has a temperature log\,$T\approx 6.2$ ($T\approx 1.6$\,MK), slightly less than the isothermal fit temperature (1.81\,MK); the hotter component has log\,$T = 6.52$ ($T=3.3$\,MK). Compared with the NA intervals, then, the X-ray emission has a DEM with two components having temperatures that are both slightly larger. A similarly improved fit to the observed spectra and their average arises from folding the DEM for the F intervals through the SphinX spectral response function. This is shown by the green histogram in Figure~\ref{fig:aver_spectrum_fla_MS}, which now closely fits the averaged spectrum of the F intervals up to at least 5\,keV. The cooler component peaks at log\,$T = 6.22$ ($T \approx 1.66$\,MK, slightly less than that of the isothermal temperature (1.86\,MK), while the hotter component is a broad feature with peak temperature of log\,$T = 6.5$ ($T \approx 3.2$\,MK). The addition of this broad feature in the DEM results in the improved fit to the spectra.

\begin{figure}
   \centerline{\includegraphics[width=12cm]{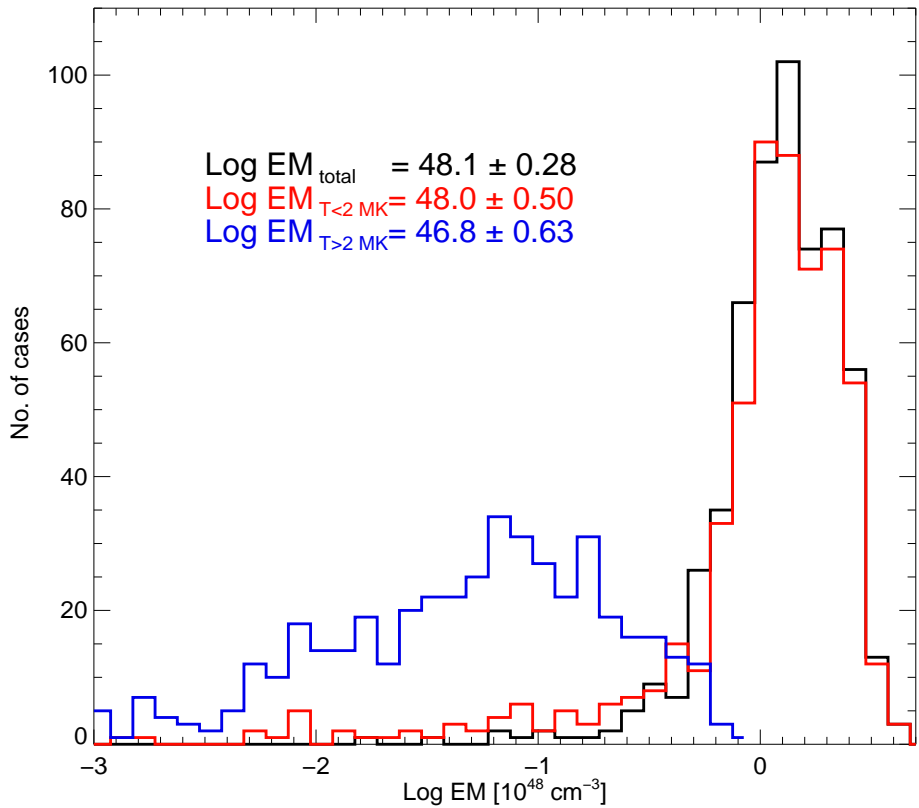}}
      \caption{The emission measure in cooler ($T < 2$\,MK, red) and hotter ($T > 2$\,MK, blue) components and total emission measure (black) are compared in the DEM analysis of 576 NA intervals. }
      \label{fig:EM_3_HISTO_from_DEM_1pt3_7_MK}
  \end{figure}

In our previous isothermal analysis of SphinX spectra obtained on 16~September \citep{jsyl12_Sph}, we identified the emission with temperature 1.7\,MK with the general coronal emission seen in coronal images in the TESIS 171\,\AA\ filter. The emission in this filter is dominated by Fe~{\sc ix} and Fe~{\sc x} lines with contribution functions peaking at about 1.0\,MK. From the SphinX emission measure ($1 \times 10^{48}$\,cm$^{-3}$) and estimated volume of this emission equal to $4 \times 10^{31}$\,cm$^3$, we obtained coronal electron densities of $1.5 \times 10^8$\,cm$^{-3}$ in agreement with other estimates. Figure~\ref{fig:EM_3_HISTO_from_DEM_1pt3_7_MK} summarizes the present analysis, showing the frequency distributions of the peak values of DEM for the hotter $T > 2$\,MK and cooler ($T < 2$\,MK) components in the DEM analyses of the 576 NA spectra. The peak of the frequency distribution for the cooler component has an emission measure that is very similar to the 16~September value ($1.4 \times 10^{48}$\,cm$^{-3}$) and with the general coronal emission volume assumed to be $4 \times 10^{31}$\,cm$^3$, we find a very similar electron density, $1.8 \times 10^{8}$\,cm$^{-3}$. As before, a unit filling factor $f = 1$ was assumed.

Our present DEM analysis thus leads to a conclusion similar to that obtained earlier \citep{jsyl12_Sph}. Evidence that the hotter ($T > 2$\,MK) DEM component is identifiable with specific features in the corona is provided by a comparison of a TESIS 171\,\AA\ image during an NA period on 3\,May at 12:20\,UT with a TESIS image in the Mg~{\sc xii} (emitted at $\approx 5$\,MK) and an image from the {\em Hinode}\/ {\em X-Ray Telescope} (XRT) with its Ti-poly filter, sensitive to relatively high temperatures (see Figure~\ref{fig:Emissivity_functions}). The highest-altitude loops of a very weak, over-the-limb active region are apparent in the images from XRT and TESIS Mg~{\sc xii} filter, taken within a few hours of each other, but not the image in the TESIS 171\,\AA\ filter. This as well as other circumstantial evidence leads us to suppose that the hotter DEM component in Figure~\ref{fig:EM_3_HISTO_from_DEM_1pt3_7_MK} is most likely due to specific coronal features.

\begin{figure}
   \centerline{\includegraphics[width=12cm]{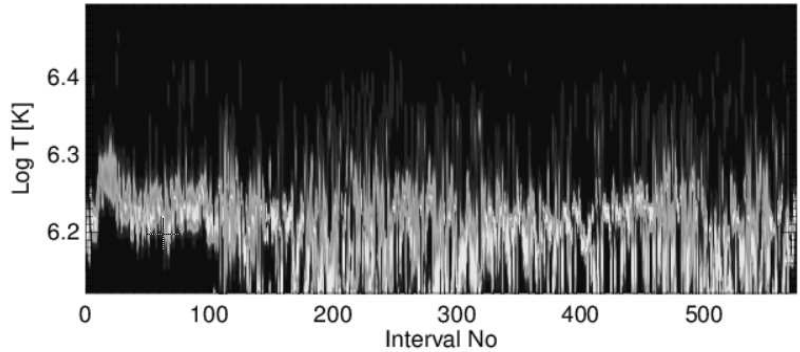}}
      \caption{Stacked DEM distributions for all 576 NA subintervals. Brighter yellow colours indicate larger values of DEM.  }
      \label{fig:Dem_stack_1pt3_7_MK}
  \end{figure}

It is instructive to plot the DEMs on a colour-coded stacked in time. This was done, based on several Fe ions in the EUV spectrum, by \cite{sch17} from measurements of the {\em Extreme-ultraviolet Variability Experiment}\/ (EVE) on the {\em Solar Dynamics Observatory}\/ for the period April 2010 to May 2014. This was from shortly after the solar minimum period when SphinX was operating to the maximum of Cycle~24 and beyond. Our time plot is shown in Figure~\ref{fig:Dem_stack_1pt3_7_MK} where, instead of time (date and UT) we plot against serial number of the 576 NA intervals, shown in Figure~\ref{fig:D1_lc}. Like the \cite{sch17} plot (Figure~5 of their paper), the cool component with log\,$T \approx 6.2$ (1.6\,MK) predominates but the hotter component with temperatures up to about log\,$T$ up to 6.4 or more ($\geqslant 2.5$\,MK) is also apparent having much smaller peak emission measure. Even for the much higher levels of solar activity seen by EVE, the cooler component is very strongly represented at all times but with higher-temperature components of DEM at periods also apparent.

\section{Synoptic Observations}
\label{sec:S-synop}

\begin{figure}
   \centerline{\includegraphics[width=12cm]{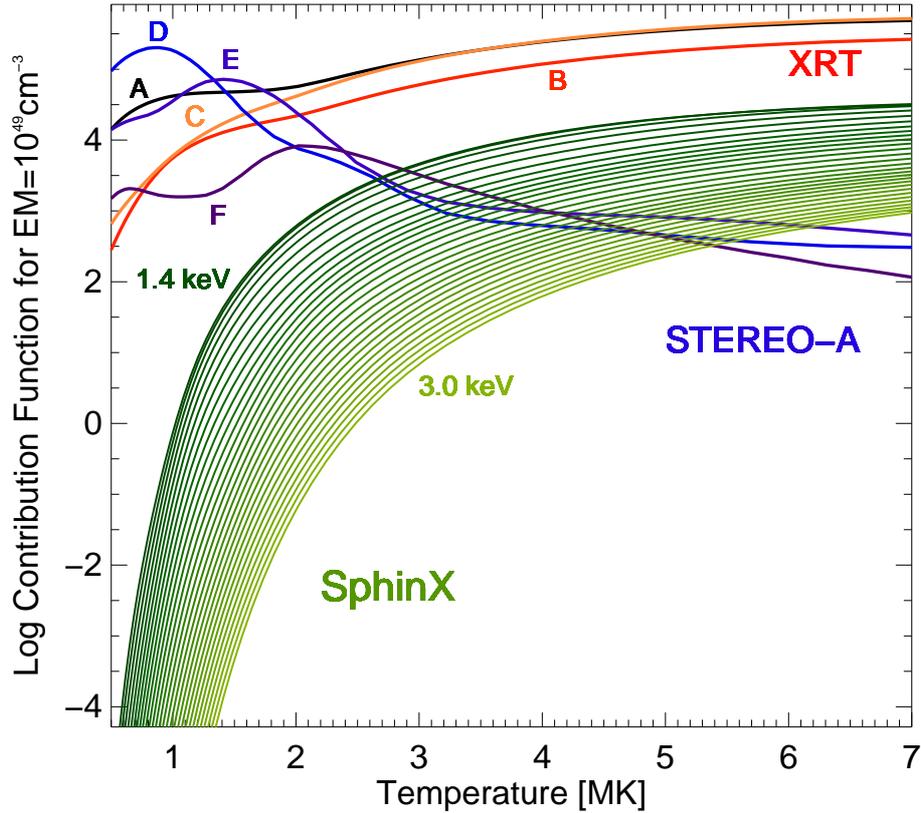}}
      \caption{Contribution functions plotted logarithmically against temperature (MK) for SphinX (green curves, energies from 1.4\,keV to 3.0\,keV in $\approx 0.05$\,keV intervals). For {\em Hinode}\/ XRT, curves are: A = black: Al-mesh; B = red: Ti-poly; C = orange: Al-poly. For STEREO-A, curves are: D = blue: 171\,\AA; E = dark blue: 195\,\AA; F = mauve: 284\,\AA. For SphinX, the contribution function expresses the sensitivity to emission in photon cm$^{-2}$\,s$^{-1}$ from a coronal plasma with an emission measure of $10^{49}$\,cm$^{-3}$. For STEREO-A and XRT, the units are DN$^{-1}$\,pixel$^{-1}$\,cm$^{-5}$\,s$^{-1}$ (DN = data number); the curves are scaled down by a factor 1000 for comparison with SphinX.}
      \label{fig:Emissivity_functions}
  \end{figure}

To check on the DEM solutions obtained from SphinX data alone, and possibly constrain the solutions better, we used synoptic observations with data available in the public domain. These include the imaging data from the XRT instrument on {\em Hinode}\/ (Al-mesh, Ti-poly, Al-poly filters) and  STEREO-A (171, 195, 284 filters). The temperature dependence of the contribution functions for SphinX detector D1 at various energies from 1.4\,keV to 3.0\,keV, shown in Figure~\ref{fig:Emissivity_functions} and obtained from on-line sources available in the SolarSoft package, is compared with those of the {\em Hinode}\/ XRT and STEREO-A channels used. For SphinX, the contribution functions express the sensitivity of each energy channel to a coronal plasma with emission measure equal to $10^{49}$\,cm$^{-3}$, while for STEREO-A and XRT, they are in units of DN$^{-1}$\,pixel$^{-1}$\,cm$^{-5}$\,s$^{-1}$ (DN = data number).

\begin{figure}
  \centerline{\includegraphics[width=9cm]{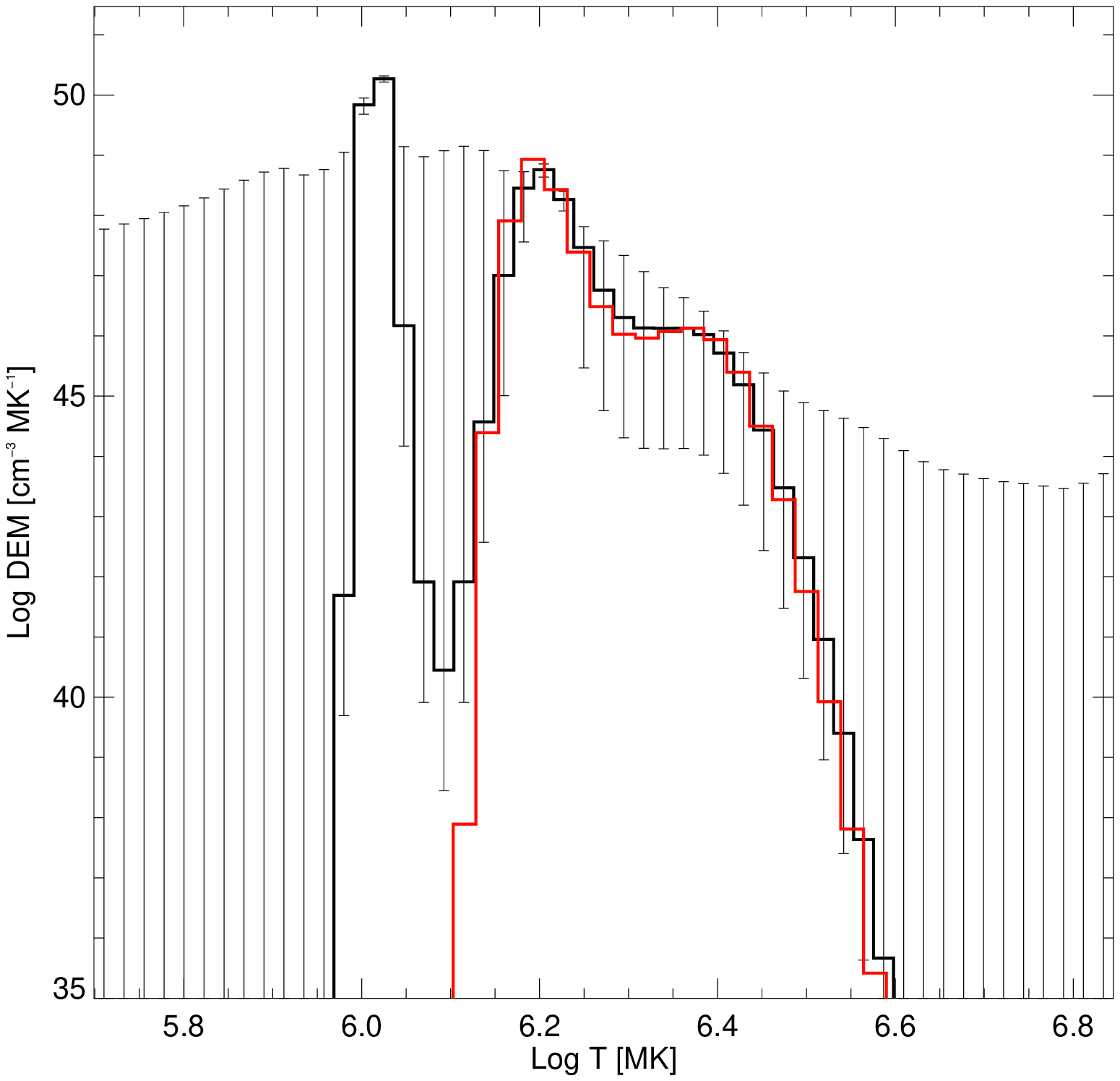}}
      \caption{The DEM solutions (plotted logarithmically) for emission in the first time interval (21:05~--~21:35\,UT on 20~February) in the 576 NA intervals. The black histogram is that derived from combining SphinX, {\em Hinode}\/ XRT, and STEREO-A data and the red histogram is that derived from SphinX data alone. Significant DEM solutions are those indicated by the histograms, while those that are indeterminate in the DEM analysis outside the range of the histograms are indicated by vertical error bars that extend below the log (DEM) = 35 level. }
      \label{fig:Dem_SphinX_XRT_Stereo}

  \end{figure}

The resulting DEM distribution for the first of the 576 NA intervals, time interval 21:05~--~21:35\,UT on 20~February, using combined SphinX, {\em Hinode}\/ XRT, and STEREO-A datasets is shown in Figure~\ref{fig:Dem_SphinX_XRT_Stereo}. On 20~February, STEREO-A was some $42^\circ$ from the Earth--Sun line, so was viewing a portion of the Sun different from {\em Hinode}\/ and {\em CORONAS-PHOTON}\/ which were in near-Earth orbits. However, we judged that the extremely low activity level of the Sun at this time, with no active regions, allowed a DEM analysis based on the combined data set. A temperature range 0.5~--~7\,MK was used for the DEM inversion. There is a clear indication of a cooler component, with well-defined peak at log\,$T = 6.0$ (black histogram in Figure~\ref{fig:Dem_SphinX_XRT_Stereo}), which is not present in the DEM solution (red histogram) obtained from SphinX data alone. We attribute this to the sensitivity of the STEREO-A 171 and 195 channels to temperatures of 1\,MK, for which the Fe~{\sc IX}/Fe~{\sc X} and Fe~{\sc XII} line emission respectively predominates - these lines have maximum emissivity at log\,$T = 6.0$ and 6.2 (see curves D and E in Figure~\ref{fig:Emissivity_functions}. The hotter component on the other hand agrees extremely well with the single component in the DEM solution obtained using SphinX data alone, including both the peak DEM and temperature. This agreement also indicates that the absolute calibrations of the SphinX and XRT (three filters) are highly consistent with each other.

\section{Discussion and Conclusions}
  \label{sec:concl}

The SphinX X-ray spectrophotometer that flew on the {\em CORONAS-PHOTON}\/ spacecraft and operated from February to November 2009 observed the total 1~--~15\,keV X-ray emission from the solar corona during the deepest solar minimum since the beginning of the Space Age. This paper reports analysis using improved data and techniques over our previous work \citep{jsyl10_Sph,jsyl12_Sph}, {\em e.g.,} in a much better determined background spectrum. We identified 576 periods of very low activity (NA), lasting from five to thirty minutes, 40 intervals when there were minor brightenings (B), and sixteen when there were micro-flares (F). Isothermal fits to the averaged NA, B, and F SphinX spectra were made, with temperatures progressively higher; these were 1.69~MK, 1.81~MK, and 1.86~MK. The emission measures for the three cases were between $0.66 \times 10^{48}$\,cm$^{-3}$ (F) and $1.17 \times 10^{48}$\,cm$^{-3}$ (NA). For the NA, B, and F averaged spectra, the isothermal fits were unsatisfactory for photon energies above about 2.5\,keV, and a higher-temperature component was needed to account for emission above this energy. Differential emission measure analyses, using the Withbroe~--~Sylwester method, were then applied to all spectra, and it was found that, in addition to a cooler component with log\,$T = 6.2$ ($T=1.6$\,MK) present for all spectra, a hotter component peaking at log\,$T = 6.38$ ($T=2.4$\,MK: NA spectra), 6.52 ($T=3.2$\,MK: B), and 6.5 ($T=3.2$\,MK: F) was present. This component has an emission measure of approximately three orders of magnitude smaller than the cooler component, but even so is needed to account for the higher-energy emission seen by SphinX. The hotter component can be identified with specific loop structures and the cooler component with general coronal emission.

Several works based on either X-ray or extreme ultraviolet spectra have arrived at varying conclusions about the presence of a hot component (up to $\approx 10$\,MK) which might be a signature of a nanoflare heating of the corona. Those of \cite{bro14} and \cite{cas15} found little evidence while that of \cite{sch09} found convincing evidence. These studies were of active regions or at least of periods when there was some low-level activity. In this work we found no evidence for a component hotter than $\approx 3$\,MK, though as the SphinX observations were based on time periods when there were no active regions this does not necessarily contradict the cited works. It is particularly noteworthy that the $0.5 - 5$\,keV X-ray spectra obtained from the rocket-borne spectrometer of \cite{cas15} are two to three orders more than the SphinX 2009 spectra from our earlier work \citep{jsyl12_Sph}. The microflare X-ray spectra described by other authors ({\em e.g.,} \cite{han08a}) are also much more energetic with higher temperatures than those described here which were exclusively unassociated with active regions, as discussed in Section~\ref{sec:Sp_anal}.

A colour-coded diagram illustrating the DEM for the NA spectra (Figure~\ref{fig:Dem_stack_1pt3_7_MK}) shows the persistence of the cooler component, and should be compared with the time series of similarly displayed spectra by \cite{sch17} over the period 2010 to 2014. Although solar activity was much higher for this period, a cool component of DEM similar in temperature to the one found from SphinX data is always present. At the higher activity levels observed by EVE, only the amount of plasma with temperatures above log\,$T = 6.2$ varies. For the solar minimum period in 2009 seen by SphinX, this higher-temperature component is still present but is much reduced.

Using imaging data from {\em Hinode}\/ XRT and STEREO-A (171, 195, 284 filters) and combining them with SphinX spectra, the DEM for the first NA period (20~February) shows a strong peak at log\,$T = 6.0$ and a hotter component peaking at around log\,$T = 6.2$ with higher-temperature shoulder. This hotter component is almost identical to that derived from SphinX spectra, but there is no trace of the cooler component. We identify the log\,$T = 6.0$ component with emission that is only evident in the STEREO-A 171 and 195 image data, this emission being primarily from the low-temperature ions Fe~{\sc IX} and Fe~{\sc X}.

It is interesting to speculate on the low level of X-ray emission for the present (2019) minimum, between Cycles 24 and 25, if observed by a SphinX-like instrument. Results from the MinXSS-2 mission, launched in late 2018 \citep{moo18}, are currently unavailable, but with its spectral resolution of 150~eV it would be interesting to compare temperatures and emission measures from this study. If the present apparent trend for ever deeper solar minima is sustained, there is clearly a need for a highly sensitive instrument to monitor the Sun's soft X-ray emission during such periods.

\begin{acks}
We acknowledge financial support from the Polish National Science centre grant number UMO-2017/25/B/ST9/01821.
\end{acks}

\section*{Disclosure of Potential Conflicts of Interest}

The authors declare that they have no conflicts of interest.


\bibliographystyle{spr-mp-sola} 
\bibliography{RESIK} 

\end{article}

\end{document}